\documentclass[12pt]{article}
\usepackage[centertags]{amsmath}
\usepackage{amsfonts}
\usepackage{amssymb}
\usepackage{amsthm}
\usepackage{newlfont}

\numberwithin{equation}{section}

\newtheorem{theorem}{Theorem}

\newtheorem{definition}{Definition}

\begin{document}
\title{Geometric quantization rules in QCPB theory}

\author{ Gen  WANG\footnote{School of Mathematical Sciences, Xiamen University,
     Xiamen, 361005, P.R.China. email:
wanggen@zjnu.edu.cn}}

\date{}

\maketitle

\begin{abstract}
Using the  quantum covariant Poisson bracket (QCPB) theory, we can accomplish much more compatible explanations of the quantum mechanics supported by the G-dynamics.   We further study the generalized quantum harmonic oscillator equipped with the G-dynamics of type I, such as geometric creation and annihilation operators, and  the geometric number operator as an extension of the number operator is well-given for the deep discussions, the geometric Hamiltonian operator is expressed as another form.  Especially, the geometric quantization rules based on the QCPB theory is then calculated.

\end{abstract}

\tableofcontents

\section{Introduction}

\subsection{Quantum harmonic oscillator}
This section mainly refers to the reference \cite{1,2,3,4,5,6}.
The quantum harmonic oscillator is described by classical Hamiltonian operator  $$\hat{H}^{\left( cl \right)}=\frac{{{\hat{p}}^{\left( cl \right)2}}}{2m}+\frac{1}{2}m{{\omega }^{2}}{{x}^{2}}$$ as one of the foundation problems of quantum mechanics, where $ {{\hat{p}}}^{{\left( cl \right)}}=-\sqrt{-1}\hbar \frac{d}{dx}$, and potential energy is $V(x)=\frac{1}{2}m{{\omega }^{2}}{{x}^{2}}$ in which $\omega$ is a constant frequency of the harmonic oscillator, in one dimension, the position of the particle was specified by a single coordinate $x$.   More precisely, it can be written as  $$\hat{H}^{\left( cl \right)}=-\frac{{{\hbar }^{2}}}{2m}\frac{{{d}^{2}} }{d{{x}^{2}}}+\frac{1}{2}m{{\omega }^{2}}{{x}^{2}}$$
The canonical commutation relation between these operators is
$\left[ x,\hat{p}^{\left( cl \right)} \right]_{QPB}=\sqrt{-1}\hbar$.
In the ladder operator method, we define the ladder operators as follows
\[a=\sqrt{\frac{m\omega }{2\hbar }}\left( x+\frac{\sqrt{-1}}{m\omega }{{\hat{p}}}^{{\left( cl \right)}}\right),~~{{a}^{\dagger }}=\sqrt{\frac{m\omega }{2\hbar }}\left( x-\frac{\sqrt{-1}}{m\omega }{{\hat{p}}}^{{\left( cl \right)}} \right)\]The operator $a$ is not Hermitian, since itself and its adjoint ${a}^{\dagger }$ are not equal.
Inverse transformation is
\[x=\sqrt{\frac{\hbar }{2m\omega }}\left( {{a}^{\dagger }}+a \right),~~{{\hat{p}}}^{{\left( cl \right)}}=\sqrt{-1}\sqrt{\frac{\hbar m\omega }{2}}\left( {{a}^{\dagger }}-a \right)\]
And it says that the ${{a}}$ and $a^{\dagger }$ operators lower and raise the energy by $\hbar \omega $ respectively. The Hamiltonian is simplified as
$$\hat{H}^{\left( cl \right)}=\hbar \omega \left( {{a}^{\dagger }}a+1/2 \right)$$ where $\left[ a,{{a}^{\dagger }} \right]_{QPB}=1$, in this ladder operator method and the eigenvalues equation is given accordingly $\hat{H}^{\left( cl \right)}\left| \psi  \right\rangle =E^{\left( cl \right)}\left| \psi  \right\rangle$,
the harmonic oscillator eigenvalues or energy levels for the mode $ \omega$ is $${{E}_{n}}^{\left( cl \right)}=\hbar \omega \left( n+1/2 \right),~n=0,1,2,\ldots$$ Then the zero point energy follows
${E}_{0}^{\left( cl \right)}=\hbar \omega/2$ at $n=0$, this paper, we denote ${E}_{0}^{\left( cl \right)}\equiv{E}_{0}=\hbar \omega/2$. This form of the frequency is the same as that for the classical simple harmonic oscillator. The most surprising difference for the quantum case is the so-called zero-point vibration of the $n=0$ ground state.

The general solution to the Schr\"{o}dinger equation leads to a sequence of evenly spaced energy levels characterized by a quantum number $n$.  Let $\xi =\sqrt{\frac{m\omega }{\hbar }}x$, then the ${{a}}$ and $a^{\dagger }$ operators can be rewritten in respect to the $\xi$ in the form
\[a=\frac{1}{\sqrt{2}}\left( \xi +\frac{d}{d\xi } \right),~~{{a}^{\dagger }}=\frac{1}{\sqrt{2}}\left( \xi -\frac{d}{d\xi } \right)\]
In the quantum harmonic oscillator, one reinterprets the ladder operators as annihilation and creation operators, adding or subtracting fixed quanta of energy to the oscillator system. Annihilation and creation operators are different for bosons (integer spin) and fermions (half-integer spin).

By using identities,
\[\xi {{\phi }_{n}}=\sqrt{\frac{n}{2}}{{\phi }_{n-1}}+\sqrt{\frac{n+1}{2}}{{\phi }_{n+1}},~~\frac{d}{d\xi }{{\phi }_{n}}=\sqrt{\frac{n}{2}}{{\phi }_{n-1}}-\sqrt{\frac{n+1}{2}}{{\phi }_{n+1}}\]
it leads to the results
\[a\left| {{\phi }_{n}} \right\rangle =\sqrt{n}\left| {{\phi }_{n-1}} \right\rangle ,{{a}^{\dagger }}\left| {{\phi }_{n}} \right\rangle =\sqrt{n+1}\left| {{\phi }_{n+1}} \right\rangle\]
It is then evident that ${a}^{\dagger}$, in essence, appends a single quantum of energy to the oscillator, while $a$ removes a quantum,  this is the reason why they are referred to as creation and annihilation operators.
It also can be rewritten as $a\left| n \right\rangle =\sqrt{n}\left| n-1 \right\rangle ,{{a}^{\dagger}}\left| n \right\rangle =\sqrt{n+1}\left| n+1 \right\rangle$ with the facts \[\left| n \right\rangle =\frac{{{\left( {{a}^{\dagger}} \right)}^{n}}}{\sqrt{n!}}\left| {~} \right\rangle ,~\left\langle m \right.\left| n \right\rangle ={{\delta }_{mn}},~\sum\limits_{n}{\left| n \right\rangle \langle n|}=1\]where $\left| {~} \right\rangle$ means the vacuum state.
Furthermore, by using the number operator $\hat N={{a}^{\dagger }}a$, it has
\begin{equation}\label{e2}
  {{\left[ \hat N,a \right]}_{QPB}}=-a,~~~{{\left[ \hat N,{{a}^{\dagger }} \right]}_{QPB}}={{a}^{\dagger }}
\end{equation}
These two relations can be easily proven.
As for annihilation and creation operators that can act on states of various types of particles, it also can refer to the ladder operators for the quantum harmonic oscillator. In the latter case, the raising operator is interpreted as a creation operator, adding a quantum of energy to the oscillator system, similarly for the lowering operator. Mathematically, the annihilation and creation operators for bosons is the same as for the ladder operators of the quantum harmonic oscillator.

\subsection{The quantum covariant Poisson bracket}
In this section, we begin with some basic concepts of quantum covariant Poisson bracket (QCPB) defined by the generalized geometric commutator \cite{7}.
In quantum mechanics, quantum Poisson bracket is the commutator of two operators $\hat{f},~\hat{g}$  defined as $\left[ \hat{f},\hat{g} \right]_{QPB}=\hat{f}\hat{g}-\hat{g}\hat{f}$.
The QCPB as an extension of  quantum Poisson bracket the is generally defined as
 \[\left[ \hat{f},\hat{g} \right]={{\left[ \hat{f},\hat{g} \right]}_{QPB}}+G\left( s,\hat{f},\hat{g} \right)\]in terms of quantum operator $\hat{f},~\hat{g}$, where $$G\left(s, \hat{f},\hat{g} \right)=\hat{f}{{\left[ s,\hat{g} \right]}_{QPB}}-\hat{g}{{\left[ s,\hat{f} \right]}_{QPB}}=-G\left(s, \hat{g},\hat{f} \right)$$ is called quantum geometric bracket, where the structural function or the geometric potential function $s$ represents the globally condition of space. Hence, 
$$ \text{\bf {QCPB=QPB+quantum geometric bracket}}$$
Note that the potential function here generally is a differentiable function. When the potential function is zero or constant, the quantum geometric brackets disappear. Clearly, zero and constant are special case.   At this time, the quantum covariant Poisson brackets (QCPB) degenerate into quantum Poisson brackets (QPB), and the corresponding results will degenerate into the results of classical quantum mechanics. Obviously, the extension of this quantum Poisson brackets is natural, mainly because the quantum theory after the extension has a wider adaptability, When $s=0$ or $s=constant$, it degenerates to the form of classical quantum theory.

It is zero if and only if $\hat{f}$ and $\hat{g}$ covariant commute, i,e. $\left[ \hat{f},\hat{g} \right]=0$.  It is remarkable to see that the QCPB representation admits a dynamical geometric bracket formula on the manifold. Note that structural function $s$ represents the background property of spacetime.

As a consequence of the precise formulation of the quantum geometric bracket, the QCPB is completely written as
\[\left[ \hat{f},\hat{g} \right]={{\left[ \hat{f},\hat{g} \right]}_{QPB}}+\hat{f}{{\left[ s,\hat{g} \right]}_{QPB}}-\hat{g}{{\left[ s,\hat{f} \right]}_{QPB}}\]Note that the last two terms provided by quantum geometric bracket can be treated as the interactions $\hat{f},\hat{g}$ with the environments.

\subsection{Covariant dynamics, generalized Heisenberg equation, G-dynamics}
In this section, we will briefly review the entire theoretical framework of quantum covariant Hamiltonian system defined by the quantum covariant Poisson bracket completely given by the previous paper.
 More precisely,
the time covariant evolution of any observable $\hat{f}$ in the covariant dynamics is given by $$\frac{\mathcal{D}\hat{f}}{dt}=\frac{1}{\sqrt{-1}\hbar }\left[ \hat{f},\hat{H} \right]$$ It contains two independent dynamics:

The generalized Heisenberg equation of motion:

$\frac{d\hat{f}}{dt}=\frac{1}{\sqrt{-1}\hbar }{{\left[ \hat{f},\hat{H} \right]}_{QPB}}-\frac{1}{\sqrt{-1}\hbar }\hat{H}{{\left[ s,\hat{f} \right]}_{QPB}}$.

The G-dynamics: $\hat{w}=\frac{1}{\sqrt{-1}\hbar }{{\left[ s,\hat{H} \right]}_{QPB}}$,\newline
Note that since quantum operators generally do not satisfy commutativity, due to  $\hat{f}\hat{w}\neq \hat{w}\hat{f}$ holds generally, it means ${{\left[ \hat{f},\hat{w} \right]}_{QPB}}\neq 0$, but here we formally record $\frac{\mathcal{D}}{dt}=\frac{d}{dt}+\hat{w}$  to represent the quantum covariant time derivative, and calculate strictly according to the derived result $\frac{\mathcal{D}}{dt}\hat{f}=\frac{d}{dt}\hat{f}+\hat{f}\hat{w}$. The specific calculation needs to be calculated according to the original definition of QCPB.
Therefore,  $\frac{\mathcal{D}}{dt}=\frac{d}{dt}+\hat{w}$ is still called covariant time operator, and
${{\hat{H}}}$ is the Hamiltonian and $[\cdot,\cdot]$ denotes the GGC of two operators.

Note that the G-dynamics is actually a frequency operator for describing the frequency of the manifold space or the environment.

Definitely, the QCPB theory leads us to a complete picture of the quantum mechanics and its deep secrets. We can definitely employ the QCPB approach to discover more hidden laws of the quantum mechanics. More importantly, the QCPB theory might  reconcile the quantum mechanics and the gravity.

Based on the G-dynamics,  the imaginary geomenergy can be completely defined, we can better give a presentation for the covariant dynamics, ect.
\begin{definition} \cite{7}
Based on the G-dynamics $\hat{w}$, then the imaginary geomenergy is induced by ${{E}^{\left( \operatorname{Im} \right)}}\left( \hat{w} \right)=\sqrt{-1}\hbar \hat{w}={{\left[ s,\hat{H} \right]}_{QPB}}$.
\end{definition}
It is obvious that the imaginary geomenergy can be expressed as
$${{E}^{\left( \operatorname{Im} \right)}}\left( \hat{w} \right)=\sqrt{-1}\hbar \hat{w}=\left[ s,\hat{H} \right]_{QPB}$$
Thus, a dynamical variable $\hat{f}$ with the geomenergy defined above, then covariant dynamics is rewritten in the form
\[\sqrt{-1}\hbar \frac{\mathcal{D}}{dt}\hat{f}=\left[ \hat{f},\hat{H} \right]=\sqrt{-1}\hbar \frac{d}{dt}\hat{f}+\hat{f}{{E}^{\left( \operatorname{Im} \right)}}\left( \hat{w} \right)\]
As a consequence of the imaginary geomenergy, we can say that imaginary geomenergy is a new kind of Hamiltonian operator.
Obviously, if the covariant equilibrium equation $\left[ \hat{f},\hat{H} \right]=0$ holds, then it yields $\frac{d}{dt}\hat{f}+\hat{f}\hat{w}=0$  or this covariant equilibrium formula in the form shows
$\frac{d}{dt}\hat{f}=-\hat{f}\hat{w}$, accordingly, $\hat{f}$ is said to be a quantum covariant conserved quantity.
It's convinced that the G-dynamics is a bridge to deeply realize the quantum mechanics.

\subsection{The QCPB for quantum harmonic oscillator}
This section mainly refers to the reference \cite{7,8}.
Commutator relations may look different than in the Schr\"{o}dinger picture, because of the time dependence of operators. For example, consider the operators $x(t)$, and ${{\hat{p}}}^{{\left( cl \right)}}(t)$. The time evolution of those operators depends on the Hamiltonian of the system. Considering the one-dimensional harmonic oscillator,
\begin{equation}\label{eq3}
  \hat{H}^{\left( cl \right)}={\frac  {{{\hat{p}}}^{{\left( cl \right)2}}}{2m}}+{\frac  {m\omega ^{{2}}x^{{2}}}{2}}
\end{equation}
the evolution of the position and momentum operators is given by:
\begin{equation}\label{eq5}
  {d \over dt}x(t)={\sqrt{-1} \over \hbar }[\hat{H}^{\left( cl \right)},x(t)]_{QPB}={\frac  {{{\hat{p}}}^{{\left( cl \right)}}}{m}}
\end{equation}\[{d \over dt}{{\hat{p}}}^{{\left( cl \right)}}(t)={\sqrt{-1}\over \hbar }[\hat{H}^{\left( cl \right)},{{\hat{p}}}^{{\left( cl \right)}}(t)]_{QPB}=-m\omega ^{{2}}x\]
As for the application of the QPB or the commutation, there are some many fields including the physics and mathematics, and so on.   As a certainly example,  we will now start by briefly reviewing the quantum mechanics of a one-dimensional harmonic
oscillator \eqref{eq3}, and see how the QCPB can be incorporated using GGC in the covariant quantization procedure.
With the Hamiltonian given by \eqref{eq3}, let's use the QCPB to recalculate the \eqref{eq5}, we can see how difference emerges. More specifically, the covariant dynamics in terms of the position reads
\[\frac{\mathcal{D}}{dt}x\left( t \right)=\frac{\sqrt{-1}}{\hbar }\left[ \hat{H}^{\left( cl \right)},x\left( t \right) \right]=\frac{{{\hat{p}}^{\left( cl \right)}}\left( t \right)}{m}+x{{\hat{w}}^{\left( cl \right)}}\]
By direct computation, the G-dynamics in terms of $\hat{H}^{\left( cl \right)}$ is given by
\begin{align}\label{eq6}
 {{\hat{w}}^{\left( cl \right)}}&=-\sqrt{-1}{{\left[ s,{{\hat{H}}^{\left( cl\right)}}\right]}_{QPB}}/\hbar=\frac{\sqrt{-1}}{\hbar }\frac{{{\hat{p}}^{\left( cl \right)2}}s+2{{\hat{p}}^{\left( cl \right)}}s{{\hat{p}}^{\left( cl \right)}}}{2m}\\
 &=b_{c}\left( 2u\frac{d}{dx}+u_{x} \right)\notag
\end{align}where $b_{c}=-\frac{\sqrt{-1}\hbar }{2m}$, and $u=\frac{ds}{dx}$, $u_{x}=\frac{{{d}^{2}}}{d{{x}^{2}}}s$ are used. Note that the G-dynamics in terms of $\hat{H}^{\left( cl \right)}$ describes the quantum rotation of the manifolds space in a certainty quantum system, its eigenvalues represent the frequency spectrum.
Furthermore, it gets $$\sqrt{-1}\hbar {{{\hat{w}}}^{\left( cl \right)}}={{\left[ s,{{\hat{H}}^{\left( cl\right)}}\right]}_{QPB}}$$
And the generalized Heisenberg equation follows
$\frac{d}{dt}x\left( t \right)=\frac{{{\hat{p}}^{\left( cl \right)}}\left( t \right)}{m}$.
In the same way,  the covariant dynamics for the momentum operator is
\begin{align}
  \frac{\mathcal{D}}{dt}{{{\hat{p}}}^{\left( cl \right)}}\left( t \right)&=\frac{\sqrt{-1}}{\hbar }\left[\hat{H}^{\left( cl \right)},{{{\hat{p}}}^{\left( cl \right)}}\left( t \right) \right]  =-m{{\omega }^{2}}x-\hat{H}^{\left( cl \right)}u+{{\hat{p}}^{\left( cl \right)}}\left( t \right)\hat{w}^{\left( cl \right)} \notag
\end{align}
where $u=\frac{ds}{dx}$ is used.
Accordingly, the generalized Heisenberg equation appears
$$\frac{d}{dt}{{{\hat{p}}}^{\left( cl \right)}}\left( t \right) =-m{{\omega }^{2}}x-\hat{H}^{\left( cl \right)}u$$

\begin{definition}[{\bf{Geomentum operator}}]\label{d5}\cite{7}
  Let $M$ be a smooth manifold represented by structural function $s$, then geomentum operator is defined as
  $$\hat{p}=-\sqrt{-1}\hbar D$$ where $D=\nabla +\nabla s$. The component is ${{\hat{p}}_{j}}=-\sqrt{-1}\hbar {{D}_{j}}$ in which ${{D}_{j}}={{\partial }_{j}}+{{\partial }_{j}}s$ holds.
\end{definition}
Note that the geomentum operator is a revision of the classical momentum operator.

\begin{theorem}[Geometric canonical quantization rules]\cite{7}
  Geometric equal-time canonical commutation relation is
  $$\left[ {{{x}_{i}}},\hat{{{p}_{j}}} \right]=\sqrt{-1}\hbar {{D}_{j}}{{x}_{i}}$$ where  $\left[ \cdot ,~\cdot  \right]={{\left[ \cdot ,~\cdot  \right]}_{QPB}}+G\left(s,\cdot ,~\cdot  \right)$ is QCPB.
\end{theorem}
Geometric canonical commutation relation can be expressed in a specific form $$\left[ {{{x}_{i}}},\hat{{{p}_{j}}} \right]=\sqrt{-1}\hbar \left( {{\delta }_{ij}}+{{x}_{i}}u_{j} \right)$$
In other words, it also can be rewritten as
$\left[ {{x}_{i}},\hat{{{p}_{j}}} \right]=\sqrt{-1}\hbar {{\theta }_{ij}}$, where
${{\theta}_{ij}}={{\delta }_{ij}}+{{x}_{i}}u_{j}$, and $u_{j}={{\partial }_{j}}s $ and ${{\partial }_{j}}=\frac{\partial }{\partial {{x}_{j}}}$.

\subsection{GEO for generalized quantum harmonic oscillator}
This section mainly refers to the reference \cite{8}.
In order to distinguish, let's rewrite the geomentum operator as $\hat{p}={{\hat{p}}^{\left( ri \right)}}=-\sqrt{-1}\hbar D$ from definition \ref{d5}, it can be alternatively used in this paper.
The covariant time evolution of operators $x(t),{{\hat{p}}}^{{\left(ri \right)}}(t)$ depends on the Hamiltonian of the system.

Considering the one-dimensional generalized quantum harmonic oscillator based upon geomentum operator \ref{d5}, the Ri-operator as a Hamiltonian operator is given by
\begin{equation}\label{eq9}
 \hat{H}^{\left( ri \right)}={\frac  {{{\hat{p}}}^{{\left(ri \right)2}}}{2m}}+{\frac  {m\omega ^{{2}}x^{{2}}}{2}}={{\hat{H}}}^{{\left(cl \right)}}-\frac{{{E}^{\left( s \right)}}}{2}-\sqrt{-1}\hbar {{\hat{w}}^{\left( cl \right)}}
\end{equation}
where $$\hat{T}^{\left( ri \right)}={\frac  {{{\hat{p}}}^{{\left(ri \right)2}}}{2m}}={{\hat{T}}^{\left( cl \right)}}-{E}^{\left( s \right)}/2-\sqrt{-1}\hbar {{\hat{w}}^{\left( cl \right)}}$$ is the geometrinetic energy operator (GEO), where

The imaginary geomenergy here:  ${{E}^{\left( \operatorname{Im} \right)}}\left( {{{\hat{w}}}^{\left( cl \right)}} \right)=\sqrt{-1}\hbar {{\hat{w}}^{\left( cl \right)}}=\frac{{{\hbar }^{2}}}{2m}\left( 2u\frac{d}{dx}+\frac{d}{dx}u \right)$,

Classical kinetic energy operator: ${{\hat{T}}^{\left( cl \right)}}=-\frac{{{\hbar }^{2}}}{2m}\frac{{{d}^{2}} }{d{{x}^{2}}}$,

The geometric potential energy function: ${{E}^{\left( s \right)}}=\frac{{{\hbar }^{2}}}{m}u^{2}$.\\
The geomentum operator in one-dimensional is ${{\hat{p}}^{\left( ri \right)}}=-\sqrt{-1}\hbar \frac{\text{D}}{dx}$, and $\frac{\text{D}}{dx}=d/dx+u$. Accordingly,
the covariant evolution of the position and  geomentum operator based on the QCHS are respectively given by:
\begin{equation}\label{eq10}
  {\mathcal{D}\over dt}x(t)={\sqrt{-1} \over \hbar }[\hat{H}^{\left(ri \right)},x(t)]
\end{equation}
\[{\mathcal{D}\over dt}{{\hat{p}}}^{{\left( ri\right)}}(t)={\sqrt{-1}\over \hbar }[\hat{H}^{\left( ri\right)},{{\hat{p}}}^{{\left( ri \right)}}(t)]\]where $[\cdot,\cdot]$ denotes the GGC of two operators.
As a result, the covariant evolution of the position is given by
$$\frac{\mathcal{D}}{dt}x\left( t \right)=\frac{{{\hat{p}}^{\left( ri \right)}}}{m}+x\hat{w}^{\left( ri \right)}$$ where G-dynamics is equal to
\begin{equation}\label{eq11}
  \hat{w}^{\left( ri \right)}=-\frac{\sqrt{-1}\hbar }{2m}\left( 2u\frac{d}{dx}+\frac{{{d}}}{d{{x}}}u+2{{u}^{2}} \right)
\end{equation}And the generalized Heisenberg equation with respect to $x$ follows
\begin{align}
 \frac{d}{dt}x\left( t \right) & =\frac{\sqrt{-1}}{\hbar }{{\left[ \hat{H}^{\left( ri \right)},x\left( t \right) \right]}_{QPB}}=\frac{{{\hat{p}}^{\left( ri \right)}}\left( t \right)}{m}\notag
\end{align}
Similarly, the covariant evolution of the geomentum operator is
\begin{align}
 \frac{\mathcal{D}}{dt}{{\hat{p}}^{\left( ri \right)}} &=-m{{\omega }^{2}}x-{{\hat{H}}^{\left( ri \right)}}u-\frac{\sqrt{-1}}{\hbar }{{\hat{p}}^{\left( ri \right)}}{{\left[ s,{{\hat{H}}^{\left( ri \right)}} \right]}_{QPB}}  \notag
\end{align}The generalized Heisenberg equation with respect to the geomentum operator ${{\hat{p}}}^{{\left( ri \right)}}$ becomes
\begin{align}
 \frac{d}{dt}{{{\hat{p}}}^{\left( ri \right)}}\left( t \right) &=-m{{\omega }^{2}}x-\hat{H}^{\left( ri \right)}u \notag
\end{align}And the G-dynamics reads
$\hat{w}^{\left( ri \right)}=-\frac{\sqrt{-1}}{\hbar }{{\left[ s,{{\hat{H}}^{\left( ri \right)}} \right]}_{QPB}}$  which is equivalent to the \eqref{eq11} as a precise expression. Accordingly, the imaginary geomenergy follows
\[{{E}^{\left( \operatorname{Im} \right)}}\left( \hat{w}^{\left( ri \right)} \right)=\sqrt{-1}\hbar \hat{w}^{\left( ri \right)}=\frac{{{\hbar }^{2}}}{2m}\left( 2u\frac{d}{dx}+\frac{{{d}}}{d{{x}}}u+2{{u}^{2}} \right)\]The G-dynamics seen from the \eqref{eq11} can evidently say how it forms and works associated with the geometric potential function.

\section{Generalized quantum harmonic oscillator}
\subsection{Ri-operator and geometric number operator}
Consider the generalized quantum harmonic oscillator \eqref{eq9},  the Ri-operator as a Hamiltonian for it is precisely written as
\begin{equation}\label{eq7}
  {{\hat{H}}^{\left( ri \right)}}=-\frac{{{\hbar }^{2}}}{2m}\left( \frac{{{d}^{2}}}{d{{x}^{2}}}+{{u}^{2}} \right)+V\left( x \right)-\sqrt{-1}\hbar {{\hat{w}}^{\left( cl \right)}}
\end{equation}
where:

The potential energy is $V\left( x \right)=\frac{1}{2}m{{\omega }^{2}}{{x}^{2}}$,

The imaginary geomenergy related to the type I is ${{E}^{\left( \operatorname{Im} \right)}}\left( {{\hat{w}}^{\left( cl \right)}} \right)=\sqrt{-1}\hbar {{\hat{w}}^{\left( cl \right)}}$.

The one-dimensional classical kinetic energy operator: ${{\hat{T}}^{\left( cl \right)}}=-\frac{{{\hbar }^{2}}}{2m}\frac{{{d}^{2}}}{d{{x}^{2}}}$,

The geometric potential energy function: ${{E}^{\left( s \right)}}=\frac{{{\hbar }^{2}}}{m}{{u}^{2}}$.\\
Notice that the first three terms $-\frac{{{\hbar }^{2}}}{2m}\left( \frac{{{d}^{2}}}{d{{x}^{2}}}+{{u}^{2}} \right)+V\left( x \right)$ are generally in a square form shown by the Ri-operator while the unique terms--the imaginary
geomenergy with respect to the G-dynamics of type I ${{E}^{\left( \operatorname{Im} \right)}}\left( {{\hat{w}}^{\left( cl \right)}} \right)=\sqrt{-1}\hbar {{\hat{w}}^{\left( cl \right)}}$ appears.
In fact, ${{E}^{\left( \operatorname{Im} \right)}}\left( {{\hat{w}}^{\left( cl \right)}} \right)$ implies that imaginary
geomenergy has a unique status, it has nothing to do with the matter, it's a structural energy form, its energy spectrum represents the deepest secrets of the space.

The geometric canonical commutation relation \cite{7} between these operator is
\[\left[ x,{\hat{p}}^{\left( ri \right)} \right]=\sqrt{-1}\hbar+\sqrt{-1}\hbar xu\]
In the ladder operator method, we define the geometric annihilation operator and geometric creation operators $${{a}^{\left( s \right)}}=a+b,~~~{{a}^{\left( s \right)}}^{\dagger }={{a}^{\dagger }}-b$$ respectively, where $b=\sqrt{\frac{\hbar }{2m\omega }}u$ is directly associated with the line curvature $u\neq0$.
By a direct computation, we can get the NG-operator given by
${{a}^{\left( g \right)}}=a-{{a}^{\dagger }}=\sqrt{\frac{2\hbar }{m\omega }}\frac{d}{dx}$
in terms of the annihilation operator $a$, similarly,
${{a}^{\left( s \right)\left( g \right)}} ={{a}^{\left( g \right)}}+2b  =\sqrt{\frac{2\hbar }{m\omega }}\frac{\text{D}}{dx}$. As a consequence of this point, we get  $${{a}^{\left( s \right)\left( g \right)}}-{{a}^{\left( g \right)}}=2b=\sqrt{\frac{2\hbar }{m\omega }}u\ne 0$$ As easily seen, the difference is only caused by geometric variable $u$ induced by the derivative of the structure function $s$ with respect to the space.
What's more, we have
\begin{align}\label{eq1}
  & {{a}^{\left( s \right)}}^{\dagger }+{{a}^{\left( s \right)}}={{a}^{\dagger }}+a \\
 & {{a}^{\left( s \right)}}^{\dagger }-{{a}^{\left( s \right)}}={{a}^{\dagger }}-a-2b \notag
\end{align}
Inverse transformation is
\[x=\sqrt{\frac{\hbar }{2m\omega }}\left( {{a}^{\left( s \right)}}^{\dagger }+{{a}^{\left( s \right)}} \right),~~{\hat{p}}^{\left( ri \right)}=\sqrt{-1}\sqrt{\frac{\hbar m\omega }{2}}\left( {{a}^{\left( s \right)}}^{\dagger }-{{a}^{\left( s \right)}} \right)\]where $${\hat{p}}^{\left( ri \right)}={{\hat{p}}^{\left( cl \right)}}-\sqrt{-1}\hbar u={{\hat{p}}^{\left( cl \right)}}-{{\left[ s,{{{\hat{p}}}^{\left( cl \right)}}\left( t \right) \right]}_{QPB}}$$ clearly is a non-Hermitian operator.  With the support of \eqref{eq1}, inverse transformation can be rewritten as
\begin{equation}\label{eq2}
  x=\sqrt{\frac{\hbar }{2m\omega }}\left( {{a}^{\dagger }}+a \right),~~{\hat{p}}^{\left( ri \right)}=\sqrt{-1}\sqrt{\frac{\hbar m\omega }{2}}\left( {{a}^{\dagger }}-a-2b \right)
\end{equation}
in terms of the ${{a}}$ and $a^{\dagger }$ operators.
Thus, we can then show that a general result for ${{a}}$ and $a^{\dagger }$ operators is given by
\[{{a}^{\left( s \right)}}^{\dagger }{{a}^{\left( s \right)}}=\frac{m\omega }{2\hbar }{{x}^{2}}-\frac{{{\hbar }}}{2{{m}}{{\omega }}}\frac{{{\text{D}}^{2}}}{d{{x}^{2}}}-1/2\]where $\frac{{{\text{D}}^{2}}}{d{{x}^{2}}}=\frac{{{d}^{2}}}{d{{x}^{2}}}
+{{u}^{2}}+{{\hat{w}}^{\left( cl \right)}}/{{b}_{c}}$, where
\begin{equation}\label{e1}
  {{\hat{w}}^{\left( cl \right)}}/{{b}_{c}}=2u\frac{d}{dx}+\frac{d u}{dx}
\end{equation}
has been used and $b_{c}=-\frac{\sqrt{-1}\hbar }{2m}$. More compactly, the geometric number operator ${{\hat{N}}^{\left( r \right)}}$ can be simply expressed as
$${{\hat{N}}^{\left( r \right)}}={{a}^{\left( s \right)}}^{\dagger }{{a}^{\left( s \right)}}=\hat{N}+{{\hat{N}}^{\left( s \right)}}$$It's obvious that this is a natural generalization of the number operator, the term ${{\hat{N}}^{\left( s \right)}}$ nicely remedies the shortage of the number operator, it helps to broaden the quantum range wider to study.
 More precisely,
\begin{align}
 {{a}^{\left( s \right)\dagger }}{{a}^{\left( s \right)}} & =\left( {{a}^{\dagger }}-b \right)\left( a+b \right)=\hat{N}+{{\hat{N}}^{\left( s \right)}}\notag \\
 & ={{a}^{\dagger }}a+{{a}^{\dagger }}b-ba-{{b}^{2}} \notag
\end{align} where
\begin{align}
  & {{\hat{N}}^{\left( s \right)}}={{a}^{\dagger }}b-ba-{{b}^{2}}=-\frac{\hbar }{2m\omega }{{\hat{w}}^{\left( cl \right)}}/{{b}_{c}}-\frac{\hbar }{2m\omega }{{u}^{2}} \notag\\
 & \hat{N}={{a}^{\dagger }}a=\frac{m\omega }{2\hbar }{{x}^{2}}-\frac{\hbar }{2m\omega }\frac{{{d}^{2}}}{d{{x}^{2}}}-1/2 \notag
\end{align}where ${b}^{2}=\frac{\hbar }{2m\omega }{{u}^{2}} $, and mixed term is expressed as
 $$ {{\hat{N}}^{\left( mix \right)}}(a,{a}^{\dagger },b)={{a}^{\dagger }}b-ba=-\frac{\hbar }{2m\omega }{{\hat{w}}^{\left( cl \right)}}/{{b}_{c}}=-\sqrt{-1}{{\hat{w}}^{\left( cl \right)}}/\omega$$ that is only linked to the G-dynamics of the type I, conversely, the term ${{a}^{\dagger }}a-{{b}^{2}}$ can be regarded as the pure term. In fact, the mixed term implies the essence of the G-dynamics which exists as a frequency operator for the frequency spectrum.  The details for ${{a}^{\dagger }}a$ can be given by
$${{a}^{\dagger }}a=\frac{m\omega }{2\hbar }\left( {{x}^{2}}+\frac{{{\hat{p}}^{\left( cl \right)2}}}{{{\left( m\omega  \right)}^{2}}}-\frac{\hbar }{m\omega } \right)=\frac{m\omega }{2\hbar }{{x}^{2}}+\frac{{{\hat{p}}^{\left( cl \right)2}}}{2\hbar m\omega }-1/2$$
Accordingly, it leads to the result
$${{\hat{N}}^{\left( s \right)}}=-\frac{\sqrt{-1}{{\hat{w}}^{\left( cl \right)}}}{\omega }-\frac{\hbar }{2m\omega }{{u}^{2}}$$Thusly, the  geometric number operator ${{\hat{N}}^{\left( r \right)}}$ can be rewritten in a clear form
\begin{equation}\label{eq8}
  {{\hat{N}}^{\left( r \right)}}={{a}^{\dagger }}a-\frac{\hbar }{2m\omega }{{u}^{2}}-\sqrt{-1}\frac{{{\hat{w}}^{\left( cl \right)}}}{\omega }
\end{equation}
The Ri-operator \eqref{eq7}  as a Hamiltonian is certainly expressed as
\begin{equation}\label{eq12}
  {{\hat{H}}^{\left( ri \right)}}=\hbar \omega \left( {{\hat{N}}^{\left( r \right)}}+1/2 \right)
\end{equation}
Plugging \eqref{eq8} into the \eqref{eq12}, it yields
\begin{align}
 {{\hat{H}}^{\left( ri \right)}} & =\hbar \omega \left( {{a}^{\left( s \right)\dagger }}{{a}^{\left( s \right)}}+1/2 \right)= \hbar \omega \left( {{a}^{\dagger }}a-{{b}^{2}}+{{a}^{\dagger }}b-ba+1/2 \right)\notag\\
 & =\hbar \omega \left( {{a}^{\dagger }}a-\frac{\sqrt{-1}{{\hat{w}}^{\left( cl \right)}}}{\omega }-\frac{\hbar }{2m\omega }{{u}^{2}}+1/2 \right) \notag\\
 & =\hbar \omega \left( {{a}^{\dagger }}a+1/2 \right)-\hbar \omega \left( \frac{\sqrt{-1}{{\hat{w}}^{\left( cl \right)}}}{\omega }+\frac{\hbar }{2m\omega }{{u}^{2}} \right) \notag\\
 & =\hbar \omega \left( {{a}^{\dagger }}a+1/2 \right)-\left( \sqrt{-1}\hbar {{\hat{w}}^{\left( cl \right)}}+\frac{{{\hbar }^{2}}}{2m}{{u}^{2}} \right) \notag\\
 & =\hbar \omega {{a}^{\dagger }}a+\hbar \omega /2-\frac{{{\hbar }^{2}}}{2m}{{u}^{2}}-\sqrt{-1}\hbar {{\hat{w}}^{\left( cl \right)}} \notag\\
 & ={{\hat{H}}^{\left( cl \right)}}+{{\hat{H}}^{\left( s \right)}} \notag
\end{align}
where the geometric Hamiltonian operator takes form
\begin{align}
 {{\hat{H}}^{\left( s \right)}} & =\hbar\omega{{\hat{N}}^{\left( s \right)}}=-\hbar \omega \left( \frac{\sqrt{-1}{{\hat{w}}^{\left( cl \right)}}}{\omega }+\frac{\hbar }{2m\omega }{{u}^{2}} \right) =\hbar \omega ({{a}^{\dagger }}b-ba-{{b}^{2}}) \notag\\
 & =-\sqrt{-1}\hbar {{\hat{w}}^{\left( cl \right)}}-\frac{{{\hbar }^{2}}}{2m}{{u}^{2}}   \notag
\end{align}
and ${{E}_{0}}=\hbar \omega /2$ is the zero point energy, and
\begin{align}
 {{\hat{H}}^{\left( cl \right)}} & =\hbar \omega \left( {{a}^{\dagger }}a+1/2 \right) \notag\\
 & =\hbar \omega \left( \frac{m\omega }{2\hbar }{{x}^{2}}+\frac{{{\hat{p}}^{\left( cl \right)2}}}{2\hbar m\omega }-1/2+1/2 \right)\notag \\
 & =\frac{{{\hat{p}}^{\left( cl \right)2}}}{2m}+\frac{m{{\omega }^{2}}}{2}{{x}^{2}} \notag
\end{align}
The eigenvalues equation is given accordingly \[{{\hat{H}}^{\left( ri \right)}}\psi =\hbar \omega \left(\hat{N}+1/2 \right)\psi +{{\hat{H}}^{\left( s \right)}}\psi =E^{\left( ri \right)}\psi \]where
\begin{align}
  {{\hat{H}}^{\left( s \right)}}\psi & =\hbar \omega {{\hat{N}}^{\left( s \right)}}\psi=-\sqrt{-1}\hbar {{\hat{w}}^{\left( cl \right)}}\psi -\frac{{{\hbar }^{2}}}{2m}{{u}^{2}}\psi={{E}^{\left(b \right)}}\psi  \notag
\end{align}
The eigenvalues or energy levels of the Ri-operator is \[{{E}_{n}}^{\left( ri \right)}=\hbar \omega \left( n+\frac{1}{2} \right)+{{E}^{\left( b \right)}},~~n=0,1,2,\ldots\]
where
${{E}^{\left( b \right)}} =-\sqrt{-1}\hbar {{w}^{\left( q \right)}}-\frac{{{\hbar }^{2}}}{2m}{{u}^{2}}$, and the energy levels are quantized at equally spaced values.

As a result given by \cite{8}, we have obtained the following result
\begin{equation}\label{eq13}
  \sqrt{-1}\hbar \left( {{\partial }_{t}}-{{\hat{w}}^{\left( cl \right)}} \right)\psi ={{\hat{H}}^{\left( ri \right)}}\psi+{{E}^{\left( s \right)}}\psi/2
\end{equation}
where ${{E}^{\left( s \right)}}=\frac{{{\hbar }^{2}}}{m}u^{2}$. Hence, inspired by the equation \eqref{eq13},   it leads to the result
$${{\hat{H}}^{\left( ri \right)}}+\frac{{{\hbar }^{2}}}{2m}{{u}^{2}}={{\hat{H}}^{\left( cl \right)}}-{{E}^{\left( \operatorname{Im} \right)}}\left( {{\hat{w}}^{\left( cl \right)}} \right)$$or in the form $${{\hat{H}}^{\left( ri \right)}}=\hbar \omega \hat{N}-\frac{{{\hbar }^{2}}}{2m}{{u}^{2}}+{{\hat{H}}^{\left(hp \right)}}$$ where

the classical Hamiltonian operator: ${{\hat{H}}^{\left( cl \right)}}=\hbar \omega \hat{N}+\hbar \omega /2$,

And ${{\hat{H}}^{\left(hp \right)}}=\hbar \omega /2-\sqrt{-1}\hbar {{\hat{w}}^{\left( cl \right)}}$,  its eigenvalues can be given by
\begin{align}\label{e3}
 {{E}^{\left( hp\right)}} &=E_{0}-\sqrt{-1}\hbar {{w}^{\left( q \right)}} =\hbar \left( \omega /2-\sqrt{-1}{{w}^{\left( q \right)}} \right)
\end{align}with respect to the eigenfunctions, it implies that a connection between the ${{E}^{\left( q \right)}}=\hbar {{w}^{\left( q \right)}}$ and zero point energy ${{E}_{0}}=\hbar \omega /2$ at zero particle $n=0$ corresponding to the vacuum condition. It reveals that the  ${{E}^{\left( q \right)}}=\hbar {{w}^{\left( q \right)}}$ describes vacuum energy of some types.
Then we get
$${{E}^{\left( hp\right)}}/\hbar =\omega /2-\sqrt{-1}{{w}^{\left( q \right)}}$$
Inspired by the zero point energy, it strongly implies that the $\sqrt{-1}{{w}^{\left( q \right)}} $ has a special significance, as a geometric frequency, it naturally links to the algebra.

\section{Geometric annihilation and creation operators}
Let's consider the geometric annihilation operator ${{a}^{\left( s \right)}}$ and geometric creation operator ${{a}^{\left( s \right)}}^{\dagger }$,
\[{{a}^{\left( s \right)}}=\frac{1}{\sqrt{2}}\left( \xi +\frac{\text{D}}{d\xi } \right),~~{{a}^{\left( s \right)}}^{\dagger }=\frac{1}{\sqrt{2}}\left( \xi -\frac{\text{D}}{d\xi } \right)\]where $\xi =\sqrt{\frac{m\omega }{\hbar }}x$.
By a direct evaluation and comparative analysis, we have $\frac{d}{d\xi }=\chi\frac{d}{dx}$, where $\chi=\sqrt{\frac{\hbar }{m\omega }}$,    then it yields $\frac{\text{D}}{d\xi }=\chi\frac{\text{D}}{dx} $, where $\frac{ds}{d\xi }=\chi u$.
Accordingly,
\begin{equation}\label{e6}
  {{a}^{\left( s \right)}}\left| {{\phi }_{n}} \right\rangle=\sqrt{n}\left| {{\phi }_{n-1}} \right\rangle+b\left| {{\phi }_{n}} \right\rangle,~{{a}^{\left( s \right)}}^{\dagger }\left| {{\phi }_{n}} \right\rangle=\sqrt{n+1}\left| {{\phi }_{n+1}} \right\rangle-b\left| {{\phi }_{n}} \right\rangle
\end{equation}
As we can see, $b{{\phi }_{n}}$ implies that it remains the numbers of particles, unlike the function of the annihilation and creation operators.
\begin{align}
  & {{a}^{\left( s \right)}}\left| n \right\rangle =\sqrt{n}\left| n-1 \right\rangle +b\left| n \right\rangle  \notag\\
 & {{a}^{\left( s \right)\dagger }}\left| n \right\rangle =\sqrt{n+1}\left| n+1 \right\rangle -b\left| n \right\rangle \notag
\end{align}
where $b=\widetilde{\chi }u={{\left( {{E}^{\left( s \right)}}/{{E}_{0}} \right)}^{1/2}}/2$.
In fact, by using the creation and annihilation operators, we can obtain the functions of both geometric annihilation and creation operators
\begin{align}
  & {{a}^{\left( s \right)}}\left| {{\phi }_{n}} \right\rangle=\left( \sqrt{n}+\frac{b}{\sqrt{n}}{{a}^{\dagger }} \right)\left| {{\phi }_{n-1}} \right\rangle =\left( 1+\frac{b}{n}{{a}^{\dagger }} \right)a\left| {{\phi }_{n}} \right\rangle\notag\\
 & {{a}^{\left( s \right)\dagger }}\left| {{\phi }_{n}} \right\rangle=\left( \sqrt{n+1}-\frac{b}{\sqrt{n+1}}a \right)\left| {{\phi }_{n+1}} \right\rangle=\left( 1-\frac{b}{n+1}a \right){{a}^{\dagger }}\left| {{\phi }_{n}} \right\rangle \notag
\end{align}
where we have used
$$\left| {{\phi }_{n}} \right\rangle=\frac{{{a}^{\dagger }}{\left| {{\phi }_{n-1}} \right\rangle}}{\sqrt{n}}=\frac{a{\left| {{\phi }_{n+1}} \right\rangle}}{\sqrt{n+1}}$$ As we can see, this is a new transformation for the annihilation and creation operators, and weird one, and
\begin{equation}\label{a1}
  a={{\widetilde{\chi }}^{-1}}x/2+\widetilde{\chi }\frac{d}{dx},~~{{a}^{\dagger }}={{\widetilde{\chi }}^{-1}}x/2-\widetilde{\chi }\frac{d}{dx}
\end{equation}
where  $\widetilde{\chi }=\frac{\chi }{\sqrt{2}}$, then $b=\widetilde{\chi }u$ appeared previously,
and then we can further get $${{\left[ s,{{a}^{\dagger }} \right]}_{QPB}}=b,~~~{{\left[ s,a \right]}_{QPB}}=-b$$ It leads to identity $${{\left[ s,{{a}^{\dagger }} \right]}_{QPB}}+{{\left[ s,a \right]}_{QPB}}=0$$
Hence, the geometric annihilation operator and geometric creation operators can be rewritten as  $${{a}^{\left( s \right)}}=a-{{\left[ s,a \right]}_{QPB}},~~~~{{a}^{\left( s \right)}}^{\dagger }={{a}^{\dagger }}-{{\left[ s,{{a}^{\dagger }} \right]}_{QPB}}={{a}^{\dagger }}+{{\left[ s,a \right]}_{QPB}}$$
Actually, \eqref{e6} can be rewritten as \[{{a}^{\left( s \right)}}\left| {{\phi }_{n}} \right\rangle=a\left| {{\phi }_{n}} \right\rangle+b\left| {{\phi }_{n}} \right\rangle,~~~~~{{a}^{\left( s \right)\dagger }}\left| {{\phi }_{n}} \right\rangle={{a}^{\dagger }}\left| {{\phi }_{n}} \right\rangle-b\left| {{\phi }_{n}} \right\rangle\]

The following geometric commutation relations can be easily obtained by substituting the geometric canonical commutation relation,
\begin{theorem}\label{t1}
The QCPB in terms of $a,{{a}^{\dagger }}$ is
$$\left[ a,{{a}^{\dagger }} \right]=1+xu$$

\begin{proof}
As classic result shows, a fact that
${{\left[ a,{{a}^{\dagger }} \right]}_{QPB}}=1$ holds,  then the QCPB reconsider it as below
$\left[ a,{{a}^{\dagger }} \right]=1+G\left(s, a,{{a}^{\dagger }} \right)$,  where more specifically, the quantum geobracket in terms of $a,{{a}^{\dagger }}$ is given by
\[G\left(s, a,{{a}^{\dagger }} \right)=a{{\left[ s,{{a}^{\dagger }} \right]}_{QPB}}-{{a}^{\dagger }}{{\left[ s,a \right]}_{QPB}}\]
By a directly computation, we respectively obtain
\[a{{\left[ s,{{a}^{\dagger }} \right]}_{QPB}}=ab\]
\[{{a}^{\dagger }}{{\left[ s,a \right]}_{QPB}}=-{{a}^{\dagger }}b\]
To combine above two parts together leads to the result of quantum geobracket
\[G\left(s, a,{{a}^{\dagger }} \right)=ab+{{a}^{\dagger }}b=ab+ba-\sqrt{-1}{{\hat{w}}^{\left( cl \right)}}/\omega\]
By using the annihilation and creation operators \eqref{a1}, given a function $\phi$, then it deduces
\begin{align}
 ab\phi +ba\phi & =\frac{{{\widetilde{\chi }}^{-1}}xb\phi }{2}+\widetilde{\chi }\frac{d\left( b\phi  \right)}{dx}+\frac{b{{\widetilde{\chi }}^{-1}}x\phi }{2}+b\widetilde{\chi }\frac{d\phi }{dx}\notag \\
 & =xu\phi +\widetilde{\chi }\frac{d\left( b\phi  \right)}{dx}+b\widetilde{\chi }\frac{d\phi }{dx}\notag \\
 & =xu\phi +\frac{\sqrt{-1}{{{\hat{w}}}^{\left( cl \right)}}\phi }{\omega } \notag
\end{align}
where \[\frac{\sqrt{-1}{{{\hat{w}}}^{\left( cl \right)}}\phi }{\omega }=\widetilde{\chi }\frac{d\left( b\phi  \right)}{dx}+b\widetilde{\chi }\frac{d\phi }{dx}=2b\widetilde{\chi }\frac{d\phi }{dx}+\phi \widetilde{\chi }\frac{db}{dx}\]
As a consequence, it gets  $G\left(s, a,{{a}^{\dagger }} \right)=xu$.  Thusly, we complete the proof as desired.
\end{proof}

\end{theorem}
With the help of theorem \ref{t1}, we are led to consider the
\[\left[ {{a}^{\left( s \right)}},{{a}^{\left( s \right)\dagger }} \right]={{\left[ {{a}^{\left( s \right)}},{{a}^{\left( s \right)\dagger }} \right]}_{QPB}}+G\left( s,{{a}^{\left( s \right)}},{{a}^{\left( s \right)\dagger }} \right)\]
In details, it has
\[\left[ a+b,{{a}^{\dagger }}-b \right]=\left[ a,{{a}^{\dagger }} \right]+\left[ b,{{a}^{\dagger }} \right]-\left[ a,b \right]\]
By simply evaluation, we prove that $\left[ b,{{a}^{\dagger }} \right]=\left[ a,b \right]$, more precisely, it can be verified
\begin{align}
  & \left[ a,b \right]={{\left[ a,b \right]}_{QPB}}+G\left( s,a,b \right)={{\left[ a,b \right]}_{QPB}}-b{{\left[ s,a \right]}_{QPB}}={{\left[ a,b \right]}_{QPB}}+{{b}^{2}} \notag\\
 & \left[ b,{{a}^{\dagger }} \right]={{\left[ b,{{a}^{\dagger }} \right]}_{QPB}}+G\left( s,b,{{a}^{\dagger }} \right)={{\left[ b,{{a}^{\dagger }} \right]}_{QPB}}+{{b}^{2}} \notag
\end{align}
and the fact
\begin{align}
 {{\left[ b,{{a}^{\dagger }} \right]}_{QPB}}-{{\left[ a,b \right]}_{QPB}} & ={{\left[ b,{{a}^{\dagger }} \right]}_{QPB}}+{{\left[ b,a \right]}_{QPB}} \notag\\
 & ={{\left[ b,{{a}^{\dagger }}+a \right]}_{QPB}} \notag\\
 & ={{\left[ b,{{\widetilde{\chi }}^{-1}}x \right]}_{QPB}}=0 \notag
\end{align}
Hence, ${{\left[ b,{{a}^{\dagger }} \right]}_{QPB}}={{\left[ a,b \right]}_{QPB}}$,
Therefore, it leads to the identity
\[\left[ {{a}^{\left( s \right)}},{{a}^{\left( s \right)\dagger }} \right]=\left[ a,{{a}^{\dagger }} \right]=1+xu\]

\subsection{Geometric commutation}

\subsubsection{Geometric commutation for number operator}

As mentioned previously, ${{a}^{\dagger }}b-ba=-\sqrt{-1}{{\hat{w}}^{\left( cl \right)}}/\omega$, where
$b=\widetilde{\chi }u$, we denote $${{\hat{N}}^{\left( cg \right)}}= {{\hat{N}}^{\left( mix \right)}}(a,{a}^{\dagger },b)={a}^{\dagger }b-ba=-\sqrt{-1}{{\hat{w}}^{\left( cl \right)}}/\omega$$ for a convenient discussions.

Using the QCPB theory, we can do more calculations such as
$$\left[\hat N,a \right]={{\left[ \hat N,a \right]}_{QPB}}+G\left( s,\hat N,a \right)$$ and
$$\left[ \hat N,{{a}^{\dagger }} \right]={{\left[\hat N,{{a}^{\dagger }} \right]}_{QPB}}+G\left( s,\hat N,{{a}^{\dagger }} \right)$$
According to \eqref{e2}, we can go farther, by calculating the quantum geometric bracket $G\left( s,\hat N,a \right)$ and $G\left( s,\hat N,{{a}^{\dagger }} \right)$, respectively. More precisely,
\begin{align}
  & G\left( s,\hat{N},a \right)=\hat{N}{{\left[ s,a \right]}_{QPB}}-a{{\left[ s,\hat{N} \right]}_{QPB}} \notag\\
 & G\left( s,\hat{N},{{a}^{\dagger }} \right)=\hat{N}{{\left[ s,{{a}^{\dagger }} \right]}_{QPB}}-{{a}^{\dagger }}{{\left[ s,\hat{N} \right]}_{QPB}} \notag
\end{align}
By making use of the formula
\begin{align}
{{\left[ s,\hat{N} \right]}_{QPB}}&={{\left[ s,{{a}^{\dagger }}a \right]}_{QPB}} \notag\\
 & ={{\left[ s,{{a}^{\dagger }} \right]}_{QPB}}a+{{a}^{\dagger }}{{\left[ s,a \right]}_{QPB}}\notag\\
 &=ba-{{a}^{\dagger }}b\notag\\
 &=\sqrt{-1}{{\hat{w}}^{\left( cl \right)}}/\omega \notag
\end{align}
As a consequence of the $a-{{a}^{\dagger }}\neq0$, then it indicates that
${{\left[ s,\hat{N} \right]}_{QPB}}\neq 0$ holds for the geometric potential function.
Then quantum geometric bracket $G\left( s,\hat{N},a \right)$ in terms of the operator $a$ is precisely expressed as
\begin{align}
 G\left( s,\hat{N},a \right) &=\hat{N}{{\left[ s,a \right]}_{QPB}}-a{{\left[ s,\hat{N} \right]}_{QPB}}\notag \\
 & =\hat{N}{{\left[ s,a \right]}_{QPB}}-a\left( {{\left[ s,{{a}^{\dagger }} \right]}_{QPB}}a+{{a}^{\dagger }}{{\left[ s,a \right]}_{QPB}} \right) \notag\\
 & =\widetilde{\chi }\left( a\left( {{a}^{\dagger }}u-ua \right)-\hat{N}u \right)\notag\\
 &=a{{\hat{N}}^{\left( cg \right)}}-\hat{N}b\notag
\end{align}
where we have used ${{\left[ s,{{a}^{\dagger }} \right]}_{QPB}}=b,~{{\left[ s,a \right]}_{QPB}}=-b$.
Meanwhile, quantum geometric bracket $G\left( s,\hat{N},{{a}^{\dagger }} \right)$ in terms of the operator ${{a}^{\dagger }}$ precisely writes
\begin{align}
 G\left( s,\hat{N},{{a}^{\dagger }} \right) &=\hat{N}{{\left[ s,{{a}^{\dagger }} \right]}_{QPB}}-{{a}^{\dagger }}{{\left[ s,\hat{N} \right]}_{QPB}} \notag\\
 & =\hat{N}{{\left[ s,{{a}^{\dagger }} \right]}_{QPB}}-{{a}^{\dagger }}\left( {{\left[ s,{{a}^{\dagger }} \right]}_{QPB}}a+{{a}^{\dagger }}{{\left[ s,a \right]}_{QPB}} \right) \notag\\
 & =\widetilde{\chi }\left( \hat{N}u+{{a}^{\dagger }}\left( {{a}^{\dagger }}u-ua \right) \right) \notag\\
 &={{a}^{\dagger }}{{\hat{N}}^{\left( cg \right)}}+\hat{N}b\notag
\end{align}
where
\[{{\left[ s,\hat{N} \right]}_{QPB}}=\sqrt{-1}\frac{{{{\hat{w}}}^{\left( cl \right)}}}{\omega }=-{{\hat{N}}^{\left( cg \right)}}\]and ${{\hat{w}}^{\left( cl \right)}}=\sqrt{-1}\omega {{\hat{N}}^{\left( cg \right)}}$.
As a result, under the QCPB theory, we can get extensive results given by
\begin{align}
  & \left[ \hat{N},a \right]=-a+a{{\hat{N}}^{\left( cg \right)}}-\hat{N}b=a\left( {{\hat{N}}^{\left( cg \right)}}-\hat{1} \right)-\hat{N}b\notag\\
 & \left[ \hat{N},{{a}^{\dagger }} \right]={{a}^{\dagger }}+{{a}^{\dagger }}{{\hat{N}}^{\left( cg \right)}}+\hat{N}b={{a}^{\dagger }}\left( {{\hat{N}}^{\left( cg \right)}} +\hat{1}\right)+\hat{N}b \notag
\end{align}
where $\hat{1}$ is unit operator. Due to ${{\hat{N}}^{\left( cg \right)}}=-\sqrt{-1}{{\hat{w}}^{\left( cl \right)}}/\omega$, plugging it into above geometric commutations, then
\begin{align}
  & \left[ \hat{N},a \right]=-a\left( \hat{1}+\sqrt{-1}{{{\hat{w}}}^{\left( cl \right)}}/\omega  \right)-\hat{N}b  \notag\\
 & \left[ \hat{N},{{a}^{\dagger }} \right]={{a}^{\dagger }}\left( \hat{1}-\sqrt{-1}{{{\hat{w}}}^{\left( cl \right)}}/\omega  \right)+\hat{N}b  \notag
\end{align}This has generalized the result given by \eqref{e2}, it reveals there are more quantum information that needs to be studied.
In conclusions, we obtain
\[\left\{ \begin{matrix}
   {{\left[ s,a \right]}_{QPB}}=-b  \\
   {{\left[ s,{{a}^{\dagger }} \right]}_{QPB}}=b  \\
   {{\left[ s,\hat{N} \right]}_{QPB}}=\sqrt{-1}{{{\hat{w}}}^{\left( cl \right)}}/\omega   \\
\end{matrix} \right.\]
By the way, we can get
\[{{\left[ s,{{{\hat{w}}}^{\left( cl \right)}} \right]}_{QPB}}={{\left[ s,-{{\gamma }_{m}}\left( \Delta s+2\nabla s\cdot \nabla  \right) \right]}_{QPB}}=2{{\gamma }_{m}}{{\left| \nabla s \right|}^{2}}=-w^{(s)}\]
where ${{\left[ s,\nabla s\cdot \nabla  \right]}_{QPB}}=-{{\left| \nabla s \right|}^{2}}$ is easily deduced. In one dimensional case, it can be rewritten as
\[{{\left[ s,{{{\hat{w}}}^{\left( cl \right)}} \right]}_{QPB}}=2{{\gamma }_{m}}{{u}^{2}}=-w^{(s)}\]where $w^{(s)}=-2{{\gamma }_{m}}{{u}^{2}}$,  and $b_{c}=-{{\gamma }_{m}}$.    As a consequence, we further obtain
\begin{align}
 {{\left[ s,{{{\hat{N}}}^{\left( cg \right)}} \right]}_{QPB}} &=-{{\left[ s,{{\left[ s,\hat{N} \right]}_{QPB}} \right]}_{QPB}}  =-\frac{\sqrt{-1}}{\omega }{{\left[ s,{{{\hat{w}}}^{\left( cl \right)}} \right]}_{QPB}} \notag\\
 & =-2\sqrt{-1}\frac{{{\gamma }_{m}}}{\omega }{{u}^{2}} \notag\\
 &=\sqrt{-1}{{w}^{\left( s \right)}}/\omega \notag
\end{align}To sum up, there are
\[{{\left[ s,\hat{N} \right]}_{QPB}}=\sqrt{-1}\frac{{{{\hat{w}}}^{\left( cl \right)}}}{\omega }=-{{\hat{N}}^{\left( cg \right)}}\]
$${{\left[ s,{{{\hat{N}}}^{\left( cg \right)}} \right]}_{QPB}}=\sqrt{-1}{{w}^{\left( s \right)}}/\omega$$

\subsubsection{Geometric commutation for geometric number operator}
Furthermore, let's take the geometric number operator into account by using the QCPB theory, for example,
\begin{align}\label{e4}
  & \left[ {{\hat{N}}^{\left( r \right)}},a \right]=\left[ \hat{N}+{{\hat{N}}^{\left( s \right)}},a \right]=\left[ \hat{N},a \right]+\left[ {{\hat{N}}^{\left( s \right)}},a \right]
  \notag
\end{align}
Actually, we only need to evaluate the following QCPB,
\begin{align}
  & \left[ {{\hat{N}}^{\left( s \right)}},a \right]={{\left[ {{\hat{N}}^{\left( s \right)}},a \right]}_{QPB}}+G\left( s,{{\hat{N}}^{\left( s \right)}},a \right) \notag
\end{align}
where
\[{{\hat{N}}^{\left( s \right)}}=-\frac{\sqrt{-1}{{\hat{w}}^{\left( cl \right)}}}{\omega }-\frac{\hbar }{2m\omega }{{u}^{2}}={{a}^{\dagger }}b-ba-{{b}^{2}}\]
or ${{\hat{N}}^{\left( s \right)}}={{\hat{N}}^{\left( cg \right)}}-{{b}^{2}}$.
In the next, by direct computations for two parts, it leads to
\[{{\left[ {{{\hat{N}}}^{\left( s \right)}},a \right]}_{QPB}}={{\left[ {{\hat{N}}^{\left( cg \right)}}-{{b}^{2}},a \right]}_{QPB}}={{\left[ {{\hat{N}}^{\left( cg \right)}},a \right]}_{QPB}}-{{\left[ {{b}^{2}},a \right]}_{QPB}}\]
and quantum geometric bracket follows
\begin{align}
 G\left( s,{{\hat{N}}^{\left( s \right)}},a \right) & ={{\hat{N}}^{\left( s \right)}}{{\left[ s,a \right]}_{QPB}}-a{{\left[ s,{{\hat{N}}^{\left( s \right)}} \right]}_{QPB}}  \notag\\
 & ={{b}^{3}}-{{\hat{N}}^{\left( cg \right)}}b-a{{\left[ s,{{\hat{N}}^{\left( cg \right)}} \right]}_{QPB}}  \notag
\end{align}
Thusly, the QCPB in terms of ${{\hat{N}}^{\left( s \right)}},a$ is equal to
\begin{align}
 \left[ {{\hat{N}}^{\left( s \right)}},a \right] & ={{\left[ {{\hat{N}}^{\left( s \right)}},a \right]}_{QPB}}+G\left( s,{{\hat{N}}^{\left( s \right)}},a \right)  \notag\\
 & ={{\left[ {{\hat{N}}^{\left( cg \right)}},a \right]}_{QPB}}-{{\left[ {{b}^{2}},a \right]}_{QPB}}-{{\hat{N}}^{\left( cg \right)}}b-a{{\left[ s,{{\hat{N}}^{\left( cg \right)}} \right]}_{QPB}}+{{b}^{3}}  \notag
\end{align}
Then, we get
\begin{align}
 \left[ {{{\hat{N}}}^{\left( r \right)}},a \right] &=a\left( {{\hat{N}}^{\left( cg \right)}}-\hat{1}-{{\left[ s,{{\hat{N}}^{\left( cg \right)}} \right]}_{QPB}} \right)  \notag\\
 & \begin{matrix}
   {} & {} & {} & {} & {}& {} & {}& {} & {} \\
\end{matrix}-\left( \hat{N}+{{\hat{N}}^{\left( cg \right)}} \right)b+{{\left[ {{{\hat{N}}}^{\left( s \right)}},a \right]}_{QPB}}+{{b}^{3}}  \notag
\end{align}
More precisely, it shows in details
\begin{align}
\left[ {{{\hat{N}}}^{\left( r \right)}},a \right] &=a\left( {{{\hat{N}}}^{\left( cg \right)}}-\hat{1}-{{\left[ s,{{{\hat{N}}}^{\left( cg \right)}} \right]}_{QPB}} \right)-\left( \hat{N}+{{{\hat{N}}}^{\left( cg \right)}} \right)b+{{\left[ {{{\hat{N}}}^{\left( s \right)}},a \right]}_{QPB}}+{{b}^{3}}  \notag\\
 & =-a\left( \hat{1}+\sqrt{-1}\frac{{{w}^{\left( s \right)}}}{\omega }+\sqrt{-1}\frac{{{{\hat{w}}}^{\left( cl \right)}}}{\omega } \right)-\left( \hat{N}-\sqrt{-1}\frac{{{{\hat{w}}}^{\left( cl \right)}}}{\omega } \right)b  \notag\\
 & \begin{matrix}
   {} & {} & {} & {}   \notag\\
\end{matrix}-{{\left[ \sqrt{-1}\frac{{{{\hat{w}}}^{\left( cl \right)}}}{\omega },a \right]}_{QPB}}-{{\left[ {{b}^{2}},a \right]}_{QPB}}+{{b}^{3}}  \notag\\
 & =-a\left( \hat{1}+\frac{\sqrt{-1}}{\omega }\left( {{w}^{\left( s \right)}}+{{{\hat{w}}}^{\left( cl \right)}} \right) \right)-\hat{N}b+\frac{\sqrt{-1}}{\omega }{{{\hat{w}}}^{\left( cl \right)}}b  \notag\\
 & \begin{matrix}
   \begin{matrix}
   {} & {} & {}  \\
\end{matrix} & {} & {} & {}   \notag\\
\end{matrix}-\frac{\sqrt{-1}}{\omega }{{\left[ {{{\hat{w}}}^{\left( cl \right)}},a \right]}_{QPB}}-{{\left[ {{b}^{2}},a \right]}_{QPB}}+{{b}^{3}}  \notag\\
 & =-a\left( \hat{1}+\frac{\sqrt{-1}{{{\hat{w}}}^{\left( ri \right)}}}{\omega } \right)-\hat{N}b+\frac{\sqrt{-1}}{\omega }{{{\hat{w}}}^{\left( cl \right)}}b-\frac{\sqrt{-1}}{\omega }{{\left[ {{{\hat{w}}}^{\left( cl \right)}},a \right]}_{QPB}}-{{\left[ {{b}^{2}},a \right]}_{QPB}}+{{b}^{3}}  \notag\\
 & =-a\left( \hat{1}+\frac{\sqrt{-1}{{{\hat{w}}}^{\left( ri \right)}}}{\omega } \right)-\hat{N}b+\frac{\sqrt{-1}}{\omega }\left( {{{\hat{w}}}^{\left( cl \right)}}b-{{\left[ {{{\hat{w}}}^{\left( cl \right)}},a \right]}_{QPB}} \right)-{{\left[ {{b}^{2}},a \right]}_{QPB}}+{{b}^{3}} \notag
\end{align}where
\begin{align}
  & {{\left[ s,\hat{N} \right]}_{QPB}}=\sqrt{-1}\frac{{{{\hat{w}}}^{\left( cl \right)}}}{\omega }=-{{{\hat{N}}}^{\left( cg \right)}} \notag\\
 & {{\left[ s,{{{\hat{N}}}^{\left( cg \right)}} \right]}_{QPB}}=\sqrt{-1}{{w}^{\left( s \right)}}/\omega  \notag\\
 & {{{\hat{N}}}^{\left( s \right)}}={{{\hat{N}}}^{\left( cg \right)}}-{{b}^{2}}\notag
\end{align}have used for the specific derivations.  \[\left[ {{{\hat{N}}}^{\left( r \right)}},a \right]=-a\left( \hat{1}+\frac{\sqrt{-1}{{{\hat{w}}}^{\left( ri \right)}}}{\omega } \right)-\hat{N}b+\frac{\sqrt{-1}}{\omega }\left( {{{\hat{w}}}^{\left( cl \right)}}b-{{\left[ {{{\hat{w}}}^{\left( cl \right)}},a \right]}_{QPB}} \right)+{{\widetilde{\chi }}^{3}}u\left( 2{{u}_{x}}+{{u}^{2}} \right)\]where \[-{{\left[ {{b}^{2}},a \right]}_{QPB}}+{{b}^{3}}=2{{\widetilde{\chi }}^{3}}u{{u}_{x}}+{{\widetilde{\chi }}^{3}}{{u}^{3}}={{\widetilde{\chi }}^{3}}u\left( 2{{u}_{x}}+{{u}^{2}} \right)\]In comparison with the equation \eqref{e2}, this formula becomes more complex, it contains the quantum information beyond our understandings, actually.

Similarly, it's a same procedure going for the ${{a}^{\dagger }}$.
$$\left[ {{\hat{N}}^{\left( r \right)}},{{a}^{\dagger }} \right]=\left[ \hat{N}+{{\hat{N}}^{\left( s \right)}},{{a}^{\dagger }} \right]=\left[ \hat{N},{{a}^{\dagger }} \right]+\left[ {{\hat{N}}^{\left( s \right)}},{{a}^{\dagger }} \right]$$ we have following result
\[{{\left[ {{{\hat{N}}}^{\left( s \right)}},{{a}^{\dagger }} \right]}_{QPB}}={{\left[ {{\hat{N}}^{\left( cg \right)}}-{{b}^{2}},{{a}^{\dagger }} \right]}_{QPB}}={{\left[ {{\hat{N}}^{\left( cg \right)}},{{a}^{\dagger }} \right]}_{QPB}}-{{\left[ {{b}^{2}},{{a}^{\dagger }} \right]}_{QPB}}\]
and
\begin{align}
 G\left( s,{{\hat{N}}^{\left( s \right)}},{{a}^{\dagger }} \right) &={{\hat{N}}^{\left( s \right)}}{{\left[ s,{{a}^{\dagger }} \right]}_{QPB}}-{{a}^{\dagger }}{{\left[ s,{{\hat{N}}^{\left( s \right)}} \right]}_{QPB}} \notag\\
 & ={{\hat{N}}^{\left( cg \right)}}b-{{a}^{\dagger }}{{\left[ s,{{\hat{N}}^{\left( cg \right)}} \right]}_{QPB}}-{{b}^{3}} \notag
\end{align}
As a result, the QCPB in terms of ${{\hat{N}}^{\left( s \right)}},{{a}^{\dagger }}$ is derived as
\[\left[ {{{\hat{N}}}^{\left( s \right)}},{{a}^{\dagger }} \right]={{\left[ {{{\hat{N}}}^{\left( s \right)}},{{a}^{\dagger }} \right]}_{QPB}}+{{\hat{N}}^{\left( cg \right)}}b-{{a}^{\dagger }}{{\left[ s,{{\hat{N}}^{\left( cg \right)}} \right]}_{QPB}}-{{b}^{3}}\]
Therefore, by direct calculations, we obtain a result for the ${{a}^{\dagger }}$,
\begin{align}
\left[ {{{\hat{N}}}^{\left( r \right)}},{{a}^{\dagger }} \right]  &=\left[ \hat{N},{{a}^{\dagger }} \right]+\left[ {{{\hat{N}}}^{\left( s \right)}},{{a}^{\dagger }} \right]={{a}^{\dagger }}\left( \hat{1}+{{\hat{N}}^{\left( cg \right)}}-{{\left[ s,{{\hat{N}}^{\left( cg\right)}} \right]}_{QPB}} \right) \notag\\
 & \begin{matrix}
   {} & {} & {}  & {} & {} & {} & {} \\
\end{matrix}+\left( \hat{N}+{{\hat{N}}^{\left( cg \right)}} \right)b+{{\left[ {{{\hat{N}}}^{\left( s \right)}},{{a}^{\dagger }} \right]}_{QPB}}-{{b}^{3}} \notag
\end{align}
Plugging all related terms into above equation for the ${{a}^{\dagger }}$, then it certainly shows
\begin{align}
 \left[ {{{\hat{N}}}^{\left( r \right)}},{{a}^{\dagger }} \right] & ={{a}^{\dagger }}\left( \hat{1}+{{{\hat{N}}}^{\left( cg \right)}}-{{\left[ s,{{{\hat{N}}}^{\left( cg \right)}} \right]}_{QPB}} \right)+\left( \hat{N}+{{{\hat{N}}}^{\left( cg \right)}} \right)b+{{\left[ {{{\hat{N}}}^{\left( s \right)}},{{a}^{\dagger }} \right]}_{QPB}}-{{b}^{3}}  \notag\\
 & ={{a}^{\dagger }}\left( \hat{1}-\frac{\sqrt{-1}}{\omega }{{{\hat{w}}}^{\left( ri \right)}} \right)+\hat{N}b-\frac{\sqrt{-1}}{\omega }\left( {{{\hat{w}}}^{\left( cl \right)}}b+{{\left[ {{{\hat{w}}}^{\left( cl \right)}},{{a}^{\dagger }} \right]}_{QPB}} \right)  \notag\\
 & \begin{matrix}
   \begin{matrix}
   {} & {} & {} & {}  \\
\end{matrix} & {} & {} & {} & {} & {}  \notag\\
\end{matrix}-{{\left[ {{b}^{2}},{{a}^{\dagger }} \right]}_{QPB}}-{{b}^{3}}  \notag
\end{align}
where we also get
\begin{align}
\left[ {{{\hat{N}}}^{\left( s \right)}},{{a}^{\dagger }} \right]  &={{\left[ {{{\hat{N}}}^{\left( cg \right)}},{{a}^{\dagger }} \right]}_{QPB}}-{{\left[ {{b}^{2}},{{a}^{\dagger }} \right]}_{QPB}}+{{{\hat{N}}}^{\left( cg \right)}}b-{{a}^{\dagger }}{{\left[ s,{{{\hat{N}}}^{\left( cg \right)}} \right]}_{QPB}}-{{b}^{3}} \notag\\
 & =-\frac{\sqrt{-1}}{\omega }{{\left[ {{{\hat{w}}}^{\left( cl \right)}},{{a}^{\dagger }} \right]}_{QPB}}-{{\left[ {{b}^{2}},{{a}^{\dagger }} \right]}_{QPB}}-\frac{\sqrt{-1}{{{\hat{w}}}^{\left( cl \right)}}b}{\omega } \notag\\
 & \begin{matrix}
   {} & {} & {} & {}  \\
\end{matrix}\begin{matrix}
   {} & {} & {}  \\
\end{matrix}+\frac{\sqrt{-1}}{\omega }{{a}^{\dagger }}{{\left[ s,{{{\hat{w}}}^{\left( cl \right)}} \right]}_{QPB}}-{{b}^{3}} \notag\\
 & =-\frac{\sqrt{-1}}{\omega }\left( {{\left[ {{{\hat{w}}}^{\left( cl \right)}},{{a}^{\dagger }} \right]}_{QPB}}+{{{\hat{w}}}^{\left( cl \right)}}b \right)-{{a}^{\dagger }}\frac{\sqrt{-1}{{w}^{\left( s \right)}}}{\omega }-{{\left[ {{b}^{2}},{{a}^{\dagger }} \right]}_{QPB}}-{{b}^{3}} \notag
\end{align}
In totally, we have obtained
\[\left[ {{{\hat{N}}}^{\left( r \right)}},a \right]=-a\left( \hat{1}+\frac{\sqrt{-1}{{{\hat{w}}}^{\left( ri \right)}}}{\omega } \right)-\hat{N}b+\frac{\sqrt{-1}}{\omega }\left( {{{\hat{w}}}^{\left( cl \right)}}b-{{\left[ {{{\hat{w}}}^{\left( cl \right)}},a \right]}_{QPB}} \right)-{{\left[ {{b}^{2}},a \right]}_{QPB}}+{{b}^{3}}\]
\[\left[ {{{\hat{N}}}^{\left( r \right)}},{{a}^{\dagger }} \right]={{a}^{\dagger }}\left( \hat{1}-\frac{\sqrt{-1}{{\hat{w}}}^{\left( ri \right)}}{\omega } \right)+\hat{N}b-\frac{\sqrt{-1}}{\omega }\left( {{{\hat{w}}}^{\left( cl \right)}}b+{{\left[ {{{\hat{w}}}^{\left( cl \right)}},{{a}^{\dagger }} \right]}_{QPB}} \right)-{{\left[ {{b}^{2}},{{a}^{\dagger }} \right]}_{QPB}}-{{b}^{3}}\]
As these valid equations expressed, we observe that the more general form of the
$\left[ {{{\hat{N}}}^{\left( r \right)}},a \right]$ and $\left[ {{{\hat{N}}}^{\left( r \right)}},{{a}^{\dagger }} \right]$ are closely associated with the G-dynamics of types I,II,III and the function $u$ shown by $b$, these geometric properties indicates the more complex characters of the quantum mechanics holds for the real reality. It also states a complex procedure to be quantized for a certain quantum case.

\section{Conclusions}
In the support of the QCPB theory, we reconsider the generalized quantum harmonic oscillator, by using the variables $u$ and the G-dynamics, we revise annihilation and creation operators to a more general form, such as geometric annihilation and geometric creation operators, especially, the geometric quantization rules based on the QCPB theory, it develops the connections between the annihilation, creation operators and the variables $u$ and the G-dynamics, it gets the useful and meaningful results.

\end{document}